\documentclass[runningheads]{llncs}
\usepackage{graphicx}

\usepackage{amssymb}

\usepackage{color}
\newcommand{\dm}[1]{\textcolor[rgb]{0,0.0,0.}{#1}}

\newcommand{\vg}[1]{\textcolor[rgb]{0.0,0.0,0.0}{#1}}

\begin{document}
\title{Can ultrasound confidence maps predict sonographers' labeling variability?}
\titlerunning{Can ultrasound confidence maps predict sonographers' labeling variability?}

\author{
Vanessa Gonzalez Duque\textsuperscript{1,2,*} \and
Leonhard Zirus\textsuperscript{1,*} \and
Yordanka Velikova\inst{1} \and
Nassir Navab\inst{1, 3} \and
Diana Mateus\inst{2}
}

\authorrunning{VG. Duque et al.}

\institute{Computer Aided Medical Procedures, Technical University of Munich, Germany. \\ 
 \and
 LS2N laboratory at Ecole Centrale Nantes, UMR CNRS 6004, Nantes, France\\
 \and
 Computer Aided Medical Procedures, John Hopkins University, Baltimore, USA. \email{vanessag.duque@tum.de} \\
 }
 
\maketitle              
{\let\thefootnote\relax\footnotetext{* Both authors share first authorship. \\
This work has been supported in part by the European Regional Development. Fund, the Pays de la Loire region on the Connect Talent scheme (MILCOM Project) and Nantes M´etropole (Convention 2017-10470),}}

\begin{abstract}
Measuring cross-sectional areas in ultrasound images is a standard tool to evaluate disease progress or treatment response.
Often addressed today with supervised deep-learning segmentation approaches, existing solutions highly depend upon the quality of experts' annotations. However, the annotation quality in ultrasound is anisotropic and position-variant due to the inherent physical imaging principles, including attenuation, shadows, and missing boundaries, commonly exacerbated with depth. This work proposes a novel approach that guides ultrasound segmentation networks to account for sonographers' 
uncertainties and generate predictions with variability similar to the experts. We claim that realistic variability can reduce overconfident predictions and improve physicians' acceptance of deep-learning cross-sectional segmentation solutions. Toward that end, we rely on a simple and efficient method to estimate Confidence Maps (CM)s from ultrasound images. The method provides certainty for each pixel for minimal computational overhead as it can be precalculated directly from the image. We show that there is a correlation between low values in the confidence maps and expert's label uncertainty. Therefore, we propose to give the confidence maps as additional information to the networks. We study the effect of the proposed use of ultrasound CMs in combination with four state-of-the-art neural networks and in two configurations: as a second input channel and as part of the loss. We evaluate our method on 3D ultrasound datasets of the thyroid and lower limb muscles. Our results show ultrasound CMs increase the Dice score, improve the Hausdorff and Average Surface Distances, and decrease the number of isolated pixel predictions. Furthermore, our findings suggest that ultrasound CMs improve the penalization of uncertain areas in the ground truth data, thereby improving problematic interpolations. 
Our code and example data will be made public at \vg{https://github.com/IFL-CAMP/Confidence-segmentation}.

\end{abstract}

\section{Introduction}

Volumetry information of muscles and organs is used to evaluate the treatment response for diseases such as hyperthyroidism~\cite{nguyen2022thyroid} or Duchenne dystrophy~\cite{pichiecchio2018muscle}. One way of measuring the volume or the evolution is by performing segmentation of the affected organs on 3D ultrasound images. This process is time-consuming and operator dependent if done manually. Moreover, expert segmentations are anisotropic and position-variant due to the inherent physical principles governing ultrasound images: sound interacts with surface layers generating areas of attenuation, shadows, missing boundaries, etc.
This paper investigates how to make deep learning segmentation models aware of the specific uncertainties the ultrasound modality introduces, which also affect experts' annotations. To this end, we rely on the concept of Confidence Maps (CMs) introduced by Kalamaris et al. ~\cite{karamalis2012ultrasound}. Based on an image-based simplified approximation of the wave propagation through the imaged mediums, CMs proposed to estimate an uncertainty value for each pixel in the image. The problem is formulated as a label propagation on a graph solved with a random walker algorithm. The resultant confidence maps (CMs) have been used until now for improving reconstruction~\cite{berge2014orientation}, registration~\cite{wein2015automatic}, and non-deep-learning bone segmentation~\cite{beitzel2012ultrasound} in ultrasound. To the best of our knowledge, we are the first to study ultrasound confidence maps in the context of Neural Networks for semantic segmentation and analyze its influence on border variability.

To address the differences in interpretation among observers in medical image segmentation, supervised learning methods often rely on ground truth data generated by popular fusion techniques such as majority voting~\cite{iglesias2015multi} or S\dm{t}aple~\cite{warfield2004simultaneous}. Nevertheless, these techniques do not capture the variations when making predictions with the model. 
Different approaches integrate segmentation uncertainty directly as part of the model’s ability to make probabilistic predictions. Baumgartner et al.~\cite{baumgartner2019phiseg}, for example, propose a hierarchical probabilistic model, while Jungo et al.~\cite{jungo2018effect} analyse the uncertainty and calibration of brain tumor segmentation using U-Net-like architectures~\cite{ronneberger2015u}. Both methods are trained with annotations from single or multiple annotators. In a similar manner, training is done for stochastic segmentation networks~\cite{monteiro2020stochastic} and Post hoc network calibration methods~\cite{rousseau2021post}. Such approaches require modifications in the architectures and extra labeling that are time-consuming and demand a hyper-parameter search. 
Instead, our first proposal consists in providing a pre-calculated Confidence Map as a second channel to a segmentation network, which does not require multiple annotations, and does not change the loss but slightly increases the number of parameters (of the input layer). 

There exist other approaches to learn label variability without modifying the network, such as label smoothing~\cite{muller2019does,islam2021spatially,lourencco2021using}, temperature scaling~\cite{guo2017calibration}, annotator error Disentangling\cite{jacob2021disentangling}, and non-parametric calibration~\cite{wenger2020non}. Most of them are applied on MRI or CT. We align ourselves with these ideas, applying our method to ultrasound, a modality characterized by blurred edges, low signal-to-noise ratio, speckle noise, and other challenges. Our second proposition to cope with these challenges is to define a cross-entropy loss based on a probabilistic ground truth called ``confidence mask", computed from the CMs and the ground truth labels. This new label is probabilistic not only in the borders, like other methods, but in the whole structure. 
Thereby, we propose to predict the ``Confidence Masks" in order to make predictions both a good segmentation but also calibrate the output probability to the confidence content.
Following our experiments, we discover that confidence masks teach the network to penalize areas with high confidence and interpolate areas with low confidence.  

\section{Methodology}
\begin{figure}[t!]
 	\centering
 	\includegraphics[width=\textwidth]{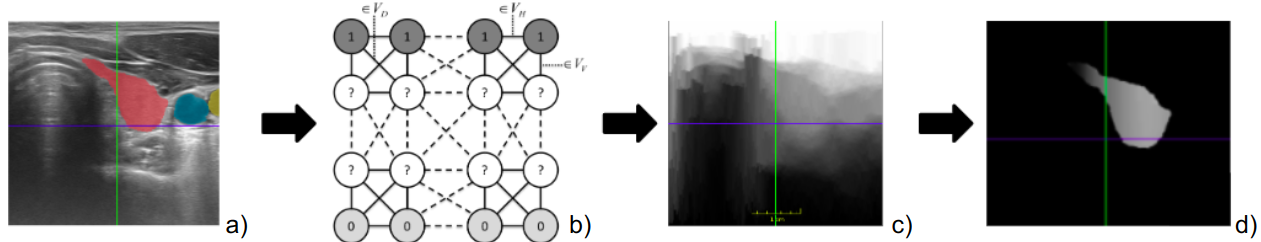}
        \caption{\dm{From confidence maps to confidence masks: a) US image and overlaid segmentations, b) image graph representation, c) Confidence map \cite{karamalis2012ultrasound}, d) Confidence mask}}
    \label{CM}
\end{figure}

\textbf{\dm{Guiding segmentation networks with Confidence Maps. }} The core of our method is to \dm{guide the training with} 
pre-calculated CMs.
\dm{Let} $\mathbf{X}\dm{\in \mathbb{R}^{W\times H\times D}}$ \dm{be} the input volume and $\mathbf{Y}, \mathbf{\hat Y}\in \mathbb{R}^{W\times H\times D\times C} $ respectively the 
one-hot encoding 
labels and \dm{the network} 
prediction
\dm{for $C$ classes.} 
\dm{We first compute the CM from the image: $\mathbf{CM}:\mathbf{X}\mapsto (0,1)^{W\times H\times D}$ (c.f. the next subsection). Our first proposition is to use the CMs as an additional channel so that the input to the network becomes $[\mathbf{X}|\mathbf{CM}]$}, with $\cdot|\cdot$ a concatenation.
\dm{In our second proposition, we combine the CMs with the labels to create a ``Confidence Mask" ($Y\cdot CM$)}, \vg{where "$\cdot$" represents the element-wise multiplication, and define the Cross entropy confidence loss over the $m$ voxels of the image as:}
\begin{equation}
 \mathrm{CE}_{conf}(\mathbf{Y}, \hat{\mathbf{Y}})=-\frac{1}{m} \sum_{i=1}^m (Y_i \cdot CM_i)\cdot\log \left(\hat{Y}_i\right)
\end{equation}

\textbf{Pre-calculated Confidence Maps}: 
In ultrasound imaging, \dm{pressure} waves are transmitted and reflected primarily to the transducers, but \dm{traversed} tissues 
absorb, diverge, refract, and disperse the sound as well. 
\dm{Therefore, as the wave progresses, the recorded}
intensities \dm{become} less reliable. The goal of the confidence map algorithm is to assign uncertainty values to each pixel in any ultrasound image, 
\dm{without prior knowledge about its content.}
Karamalis et al.\cite{karamalis2012ultrasound} proposed a simplified but efficient wave propagation model 
\dm{to address this problem}
based on a random walk on a graph.
The nodes of the graph represent the image pixels while an 8 neighbourhood rule is used \dm{to define} the edges. Edge weights model \dm{ultrasound} 
physical properties: an exponential Beer–Lambert attenuation \dm{governed by parameter $\alpha$ in the vertical direction; a penalization for horizontal and diagonal propagations associated with the beam shape; and a penalization between neighbour pixels with different intensities, controlled by parameter $\beta$, modeling 
the negative correlations between reflection and transmission across tissue boundaries. 
}
At each step of the random walk, the probability of moving from one pixel to another is based on
\dm{the defined edge weights.}   
\dm{By definition, source and sink nodes are placed at the top and bottom of the image, respectively.  
The problem is then formulated as computing the probability of the random walk starting from a transducer/source pixel to reach a sink node (c.f. Fig.~\ref{CM}-c). The sought probabilities are obtained by solving a linear system of equations, we refer the reader to \cite{karamalis2012ultrasound} for more details.
In practice, and following the above model,
the random walk goes from the top to the 
bottom approximately perpendicular to the beam/scanline direction. Deviations in the horizontal/diagonal directions are possible to a small degree, according to the image content, and controlled by the $\mathbf{\alpha}$ and $\beta$ hyper-parameters.
}\\

\begin{figure}[h]
 	\centering
 	\includegraphics[scale=0.30]{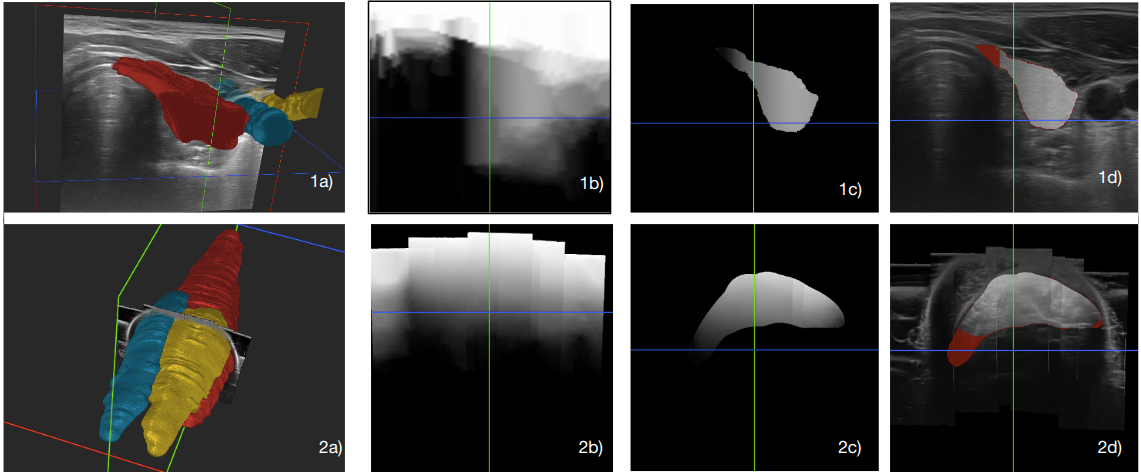}
    \caption{ Examples of the thyroid (top row) and the low-limb muscles (bottom row), respectively. 
     (a) corresponds to the 3D view of the labels at the top, the red, blue, and yellow correspond to the Thyroid, the carotid artery, and the Jugular vein, while at the bottom, they correspond to the Soleus, the Gastrocnemius lateralis, and the Gastrocnemius Medialis. (b) the CM cross-sectional view, (c) the confidence Mask used for the loss, (d) the CM overlapped over the image with red signalizing the areas with low confidence.}
    \label{dataset}
\end{figure}

\textbf{Datasets:}
Two different 3D ultrasound datasets were used for the experiments and are presented in Fig.~\ref{dataset}. 
\vg{They consist of 2D B-mode ultrasound sweeps that can generate compounded 3D volumes. }
The first dataset is open-source and available from \cite{kronke2022tracked}. It contains scans of the \textit{thyroid} of 16 volunteers, with 3 labels: thyroid, aorta, and jugular vein. Each volume contains around ~200 images of size $W\times H = 400 \times 270$, for a total of more than 1600 images.
The second in-house dataset~\cite{crouzier2018neuromechanical} contains 4 to 6 scans per volume of the left \textit{low-limb legs} of 16 participants with 3 labels: Soleus, Gastrocnemius lateralis, and Gastrocnemius Medialis muscles. Each volume of size \dm{$W\times H = 500 \times 420$} contains around ~1500 images, for a total of more than 24000 images.

For both datasets, confidence maps were calculated over 2D ultrasound images using the implementation of the random walker algorithm 
\cite{karamalis2012ultrasound} 
available in the ImFusion\footnote{ImFusion GmbH, Munich, Germany}\cite{wein2015automatic} software, version 2.36.3, with $\alpha$ and $\beta$ parameters set to $0.5, 100$ respectively. The data was split patient-wise into 11 train volumes, 2 for validation and 3 for testing, \dm{in each case}.\\

\textbf{Evaluation Metrics:}
For multi-label segmentation, we compute 8 different metrics: Dice Similarity Coefficient (DSC), 
mean Intersection over Union (mIoU), precision, recall, Average Surface Distance (ASD), Hausdorff distance(HD), miss rate, and fall out.
\dm{Following ~\cite{reinke2021common}, we evaluate the metrics for each class 
and average them over the organs and participants.
}

\section{Experiments and Results}
\textbf{Contribution of the Confidence Maps:} 
\dm{We denote the models relying on the CMs}
 as a second channel 
 \dm{with names including the term} (*-2ch-*) 
 \dm{and those with CMs in the loss}
 \dm{with the pattern} (*-*-*conf). 
 \dm{Based on a U-Net architecture we evaluate}
 a total of 10 configurations: 
 \begin{itemize}
     \item \textbf{Baselines:}
     unet-1ch-dice, unet-1ch-crossentropy(CE), unet-1ch- Dice cross entropy (diceCE)
     \item \textbf{CMs as 2nd channel:}
     unet-2ch-dice, unet-2ch-CE, unet-2ch-diceCE,
     \item \textbf{CMs within the loss :}
     unet-1ch-CEconfidence, unet-1ch-diceCEconfidence
     \item \textbf{CMS both as 2nd channel and within the loss:}
     unet-2ch-diceCE confidence and unet-2ch-CEconfidence
 \end{itemize}

\begin{figure}[h!]
 	\centering
        \includegraphics[scale=0.48]{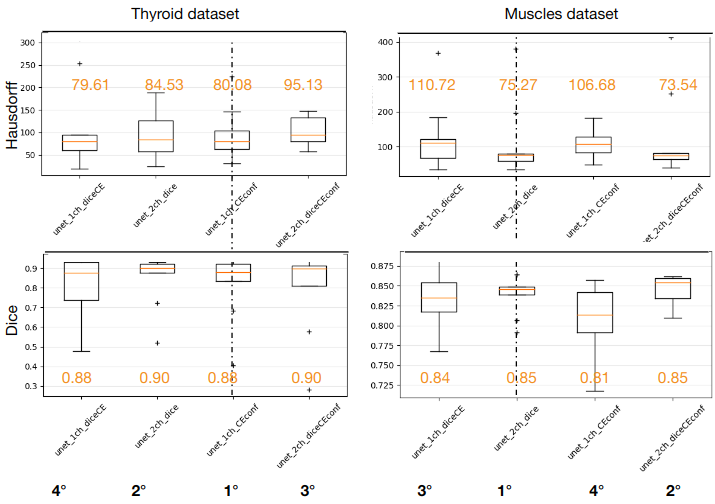}
        \caption{First and second columns report the results for the thyroid and the muscles metrics, respectively. 
        The best four performing methods are ranked 1$^\circ$, 2$^\circ$, 3$^\circ$, 4$^\circ$.
        }
    \label{UnetConfigs}
\end{figure} 

\begin{figure}[h!]
    \centering
    \includegraphics[scale=0.42]{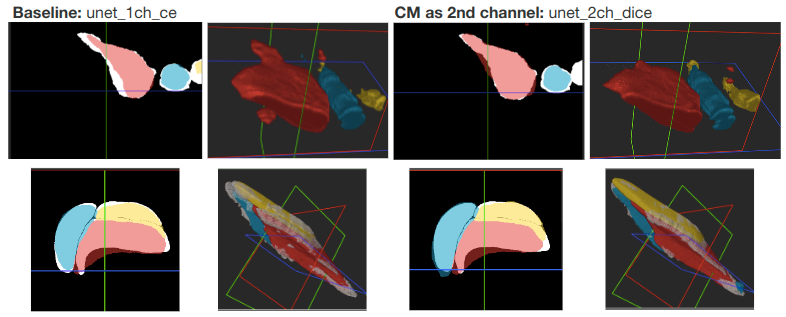}
    \caption{Prediction for one participant of the thyroid dataset in the top and the leg in the bottom. At left the baseline method:\textbf{unet-1ch-ce}, at the right our proposal:\textbf{unet-2ch-dice} }
    \label{UnetConfigs2}
\end{figure}

\dm{The results are reported in Fig. \ref{UnetConfigs}, where we keep the best configuration for each group.} 
We computed a 3-fold-cross validation to verify the independence of 
\dm{the results to the participants split.}
We found that CMs decrease the standard deviation \dm{of the DSC in general.}
While the HD of CM configurations is similar or increased for the thyroid dataset, the positive effect of CMs in the muscles dataset is clear. We attribute this behavior to the more complex muscle shapes. 
\vg{More boxplots metrics can be found in our github.}
Based on the balance between DSC and HD scores, the best two configurations are: unet-2ch-dice and unet-1ch-CEconfidence. 
\vg{Figure \ref{UnetConfigs2} showcases segmentation improvements using CMs. For the thyroid dataset, CMs reduced isolated regions, enhancing accuracy. For the leg dataset, CMs improved interpolation and smoothness of segmented structures.}

\begin{figure}[t!]
 	\centering
        \includegraphics[width = \textwidth]{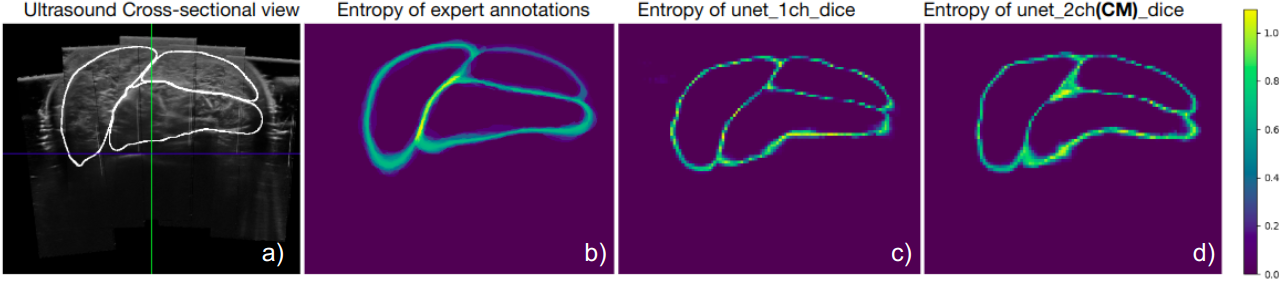}
    \caption{
    Labelling variability: 
    Ultrasound image with labeling of the muscles, b) Expert variability, c) Monte Carlo dropout for the baseline method. d) Monte Carlo dropout of our method with CMs as the second channel.}
    \label{MonteCarlo}
\end{figure}

\textbf{Expert uncertainty and prediction variability : }
To evaluate the areas where the network is less certain,
we ask the same expert to perform the labeling of the same image 100 times at different times. We compare the variability with the entropy of 100 Monte Carlo Dropout predictions for the unet-1ch-dice and unet-2ch-dice.
We observe in Fig.~\ref{MonteCarlo}, how CMs bring the predictions \dm{variability closer to the expert's uncertainty, with an anisotropic behaviour that reflects difficult areas (the intersection of the three muscles) and increases with depth.}

\begin{figure}[h!]
 	\centering
 	\includegraphics[scale=0.29]{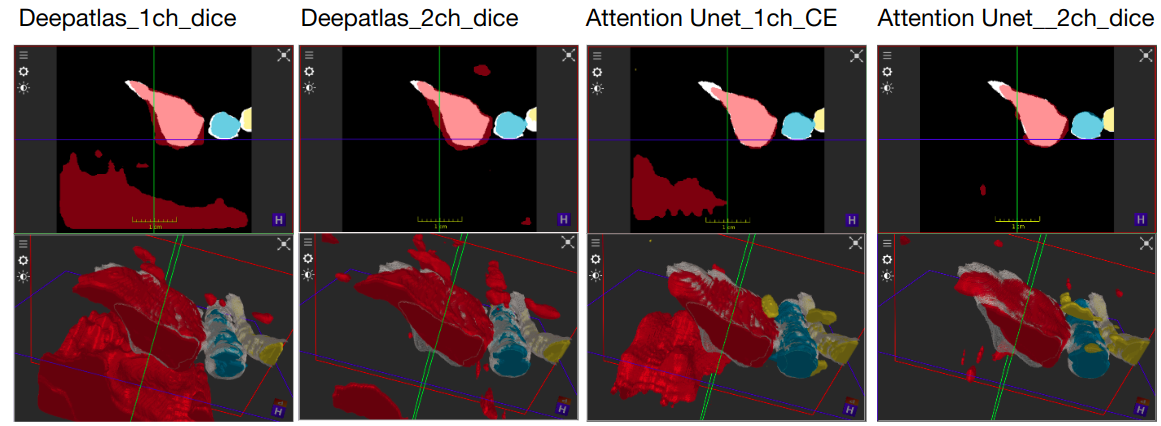}
    \caption{
    \dm{\textbf{(top row)} 2D cross-sectional view of the predictions overlaid over the expert's labels. \textbf{(bottom row)} 3D predictions in color and labels in gray for the 2 best configurations and their corresponding baselines: 1. Deep-atlas 1ch dice, 2. Deep-atlas 2ch dice, 3. Attention Unet 1ch CE, 4.Attention Unet-2ch-Dice.  }
    }
    \label{networkImages}
\end{figure}

\textbf{CMs with state of the art architectures:}
For both datasets, we tested four different 3D networks
available in MONAI\cite{cardoso2022monai}. 
We trained the models for 120 epochs, with a learning rate of 0.001 and the ADAM optimizer, on a Nvidia 390 GPU. 
Qualitative images are presented in Fig.~\ref{networkImages}. The evaluation metrics for the Muscles dataset, computed in a 3-fold cross-validation manner, are presented in Table \ref{architectures}. 

Unet transformer (UNETR) presents a very low accuracy probably as they need more data. Instead, Attention Unet obtains a very high accuracy using confidence maps, we attribute this behavior to
to the way the attention layers learn from the new meaningful additional information in the CMs.

\begin{table}[h!]
    \centering
    \caption{Metrics on the muscle dataset for the baselines and the modified versions of 4 different networks: UNet3D\cite{ronneberger2015u}, attention Unet (AttUNet) \cite{oktay2018attention}, DeepAtlas\cite{xu2019deepatlas} and UNet Transformer(UNETR)\cite{hatamizadeh2022unetr}.}
    \begin{tabular}{c|c|c|c|c}
    \hline\hline
        Network & DSC $\uparrow$ & precision $\uparrow$ & miss rate $\downarrow$ & ASD $\downarrow$  \\
        \noalign{\smallskip}\hline\noalign{\smallskip}
        UNet 1ch dice CE & $0.84\pm 0.02$ & $0.84\pm 0.06$ & $0.16\pm 0.04$ & $8.37\pm 2.19$ \\
        UNet 2ch dice$*$ & $\mathbf{0.85 \pm 0.01}$ & $0.85 \pm 0.01$ & $\mathbf{0.16 \pm 0.02}$ & $\mathbf{8.20 \pm 1.81}$ \\
        UNet 1ch diceCEconf$*$ & $0.81\pm 0.01$ & $\mathbf{0.86\pm 0.05}$ & $0.18\pm 0.04$ & $8.36\pm 0.78$ \\
        \noalign{\smallskip}\hline\noalign{\smallskip}
        AttUNet 1ch CE & $0.57 \pm 0.49$ & $0.58 \pm 0.50$ & $0.44 \pm 0.49$ & $40.00 \pm 15.01$ \\
        AttUNet 2ch dice $*$ & $\mathbf{0.86 \pm 0.00}$ & $\mathbf{0.85 \pm 0.04}$ & $\mathbf{0.12 \pm 0.04}$ & $\mathbf{7.03 \pm 1.08}$ \\
        AttUNet 1ch diceCEconf & $0.57 \pm 0.50$ & $0.59 \pm 0.51$ & $0.44 \pm 0.48$ & $40.00 \pm 14.09$ \\
        \noalign{\smallskip}\hline\noalign{\smallskip}
        DeepAtlas Dice & $0.82 \pm 0.02$ & $0.84 \pm 0.05$ & $0.18 \pm 0.06$ & $\mathbf{11.22 \pm 1.54}$ \\
        DeepAtlas 2ch dice $*$ & $0.84 \pm 0.01$ & $0.84 \pm 0.06$ & $0.15 \pm 0.07$ & $11.89 \pm 6.00$ \\
        DeepAtlas 1ch diceCEconf $*$ & $\mathbf{0.85 \pm 0.01}$ & $\mathbf{0.84 \pm 0.02}$ & $\mathbf{0.13 \pm 0.02}$ & $11.74 \pm 5.50$ \\
        \noalign{\smallskip}\hline\noalign{\smallskip}
        UNETR 1ch CE & $\mathbf{0.66\pm 0.06}$ & $0.72 \pm 0.12$ & $\mathbf{0.37 \pm 0.02}$ & $23.52 \pm 8.97$ \\
        UNETR 2ch dice$*$ & $0.48 \pm 0.12$ & $0.77 \pm 0.11$ & $0.62 \pm 0.12$ & $23.74 \pm 8.72$ \\
        UNETR 1ch diceCEconf$*$ & $0.56 \pm 0.07$ & $\mathbf{0.82 \pm 0.13}$ & $0.55 \pm 0.09$ & $\mathbf{19.54 \pm 9.65}$ \\
        \hline\hline\noalign{\smallskip}
    \end{tabular}
    \label{architectures}
\end{table}
 
We performed the Friedman statistical significance test for the Null hypothesis: ``all methods perform equally". A pair-wise post-hoc cross-validation between the baseline networks and the modified networks (2ch and dice-CEconf) was performed. The methods that reject the null hypothesis, $p<0.05$, for the Hausdorff distance, are marked with $*$.
\dm{The results in the table show that CMs} 
improve segmentation metrics by a small factor. \dm{However, looking at the qualitative results, we see CM-models}
favor better interpolation of the bottom areas where uncertainty is higher, improve the segmentation of small structures, and decrease the number of islands, as it can be seen in Fig.~\ref{networkImages}.

\section{Discussion and Conclusions}
In conclusion, this study presents an original approach to improve the awareness of ultrasound deep learning segmentation methods to label variability.
Our method, based on ultrasound Confidence Maps, takes into account  \dm{the basic ultrasound wave propagation principles, which affect}
sonographers' uncertainty when annotating. Introduced as an additional input channel or within the loss, CMs guide the network 
\dm{to predict segmentations that
effectively reproduce expert-like borders variability and whose drop-out uncertainty grows with the depth, as expected for ultrasound images. Thereby, our method can be used to generate multiple solutions for the physicians to judge, with fixed borders for certain and variable mask predictions for uncertain regions.} 
Two advantages of the approach with the CM loss is that it does not increase the number of parameters, which indicates that it is architecture agnostic. In this sense, the approach could be applied as a fine-tuning strategy after transfer learning. 

Our experimental results show that the training \dm{of CM models does not affect the convergence for either of the proposed approaches}. 
\dm{Moreover, the CMs pre-computation is very fast as it consists of the resolution of a linear system with a sparse matrix. }
In sum, this novel, simple and effective approach to introduce ultrasound and expert knowledge

can be easily implemented in combination with various ultrasound segmentation architectures without incurring additional computational costs. We evaluated our method on two datasets, one private and another public, to ensure repeatability. Future work aims at distilling the confidence map automatically. Although we used the dice loss and cross-entropy loss, other losses or combinations could also be considered.

\bibliographystyle{splncs04}
\bibliography{bibOfficial}

\begin{thebibliography}{10}
\providecommand{\url}[1]{\texttt{#1}}
\providecommand{\urlprefix}{URL }
\providecommand{\doi}[1]{https://doi.org/#1}

\bibitem{baumgartner2019phiseg}
Baumgartner, C.F., Tezcan, K.C., Chaitanya, K., H{\"o}tker, A.M., Muehlematter,
  U.J., Schawkat, K., Becker, A.S., Donati, O., Konukoglu, E.: Phiseg:
  Capturing uncertainty in medical image segmentation. In: Medical Image
  Computing and Computer Assisted Intervention--MICCAI 2019: 22nd International
  Conference, Shenzhen, China, October 13--17, 2019, Proceedings, Part II 22.
  pp. 119--127. Springer (2019)

\bibitem{beitzel2012ultrasound}
Beitzel, J., Ahmadi, S.A., Karamalis, A., Wein, W., Navab, N.: Ultrasound bone
  detection using patient-specific ct prior. In: 2012 Annual International
  Conference of the IEEE Engineering in Medicine and Biology Society. pp.
  2664--2667. IEEE (2012)

\bibitem{berge2014orientation}
Berge, C.S.z., Kapoor, A., Navab, N.: Orientation-driven ultrasound compounding
  using uncertainty information. In: Information Processing in
  Computer-Assisted Interventions: 5th International Conference, IPCAI 2014,
  Fukuoka, Japan, June 28, 2014. Proceedings 5. pp. 236--245. Springer (2014)

\bibitem{cardoso2022monai}
Cardoso, M.J., Li, W., Brown, R., Ma, N., Kerfoot, E., Wang, Y., Murrey, B.,
  Myronenko, A., Zhao, C., Yang, D., et~al.: Monai: An open-source framework
  for deep learning in healthcare. arXiv preprint arXiv:2211.02701  (2022)

\bibitem{crouzier2018neuromechanical}
Crouzier, M., Lacourpaille, L., Nordez, A., Tucker, K., Hug, F.:
  Neuromechanical coupling within the human triceps surae and its consequence
  on individual force-sharing strategies. Journal of Experimental Biology
  \textbf{221}(21) (2018)

\bibitem{guo2017calibration}
Guo, C., Pleiss, G., Sun, Y., Weinberger, K.Q.: On calibration of modern neural
  networks. In: International conference on machine learning. pp. 1321--1330.
  PMLR (2017)

\bibitem{hatamizadeh2022unetr}
Hatamizadeh, A., Tang, Y., Nath, V., Yang, D., Myronenko, A., Landman, B.,
  Roth, H.R., Xu, D.: Unetr: Transformers for 3d medical image segmentation.
  In: Proceedings of the IEEE/CVF winter conference on applications of computer
  vision. pp. 574--584 (2022)

\bibitem{iglesias2015multi}
Iglesias, J.E., Sabuncu, M.R.: Multi-atlas segmentation of biomedical images: a
  survey. Medical image analysis  \textbf{24}(1),  205--219 (2015)

\bibitem{islam2021spatially}
Islam, M., Glocker, B.: Spatially varying label smoothing: Capturing
  uncertainty from expert annotations. In: Information Processing in Medical
  Imaging: 27th International Conference, IPMI 2021, Virtual Event, June
  28--June 30, 2021, Proceedings 27. pp. 677--688. Springer (2021)

\bibitem{jacob2021disentangling}
Jacob, J., Ciccarelli, O., Barkhof, F., Alexander, D.C.: Disentangling human
  error from the ground truth in segmentation of medical images. ACL (2021)

\bibitem{jungo2018effect}
Jungo, A., Meier, R., Ermis, E., Blatti-Moreno, M., Herrmann, E., Wiest, R.,
  Reyes, M.: On the effect of inter-observer variability for a reliable
  estimation of uncertainty of medical image segmentation. In: Medical Image
  Computing and Computer Assisted Intervention--MICCAI 2018: 21st International
  Conference, Granada, Spain, September 16-20, 2018, Proceedings, Part I. pp.
  682--690. Springer (2018)

\bibitem{karamalis2012ultrasound}
Karamalis, A., Wein, W., Klein, T., Navab, N.: Ultrasound confidence maps using
  random walks. Medical image analysis  \textbf{16}(6),  1101--1112 (2012)

\bibitem{kronke2022tracked}
Kr{\"o}nke, M., Eilers, C., Dimova, D., K{\"o}hler, M., Buschner, G.,
  Schweiger, L., Konstantinidou, L., Makowski, M., Nagarajah, J., Navab, N.,
  et~al.: Tracked 3d ultrasound and deep neural network-based thyroid
  segmentation reduce interobserver variability in thyroid volumetry. Plos one
  \textbf{17}(7),  e0268550 (2022)

\bibitem{lourencco2021using}
Louren{\c{c}}o-Silva, J., Oliveira, A.L.: Using soft labels to model
  uncertainty in medical image segmentation. In: International MICCAI
  Brainlesion Workshop. pp. 585--596. Springer (2021)

\bibitem{monteiro2020stochastic}
Monteiro, M., Le~Folgoc, L., Coelho~de Castro, D., Pawlowski, N., Marques, B.,
  Kamnitsas, K., van~der Wilk, M., Glocker, B.: Stochastic segmentation
  networks: Modelling spatially correlated aleatoric uncertainty. Advances in
  neural information processing systems  \textbf{33},  12756--12767 (2020)

\bibitem{muller2019does}
M{\"u}ller, R., Kornblith, S., Hinton, G.E.: When does label smoothing help?
  Advances in neural information processing systems  \textbf{32} (2019)

\bibitem{nguyen2022thyroid}
Nguyen, D.T., Choi, J., Park, K.R.: Thyroid nodule segmentation in ultrasound
  image based on information fusion of suggestion and enhancement networks.
  Mathematics  \textbf{10}(19), ~3484 (2022)

\bibitem{oktay2018attention}
Oktay, O., Schlemper, J., Folgoc, L.L., Lee, M., Heinrich, M., Misawa, K.,
  Mori, K., McDonagh, S., Hammerla, N.Y., Kainz, B., et~al.: Attention u-net:
  Learning where to look for the pancreas. IMIDL Conference (2018)  (2018)

\bibitem{pichiecchio2018muscle}
Pichiecchio, A., Alessandrino, F., Bortolotto, C., Cerica, A., Rosti, C.,
  Raciti, M.V., Rossi, M., Berardinelli, A., Baranello, G., Bastianello, S.,
  et~al.: Muscle ultrasound elastography and mri in preschool children with
  duchenne muscular dystrophy. Neuromuscular Disorders  \textbf{28}(6),
  476--483 (2018)

\bibitem{reinke2021common}
Reinke, A., Tizabi, M.D., Sudre, C.H., Eisenmann, M., R{\"a}dsch, T.,
  Baumgartner, M., Acion, L., Antonelli, M., Arbel, T., Bakas, S., et~al.:
  Common limitations of image processing metrics: A picture story. arXiv
  preprint arXiv:2104.05642  (2021)

\bibitem{ronneberger2015u}
Ronneberger, O., Fischer, P., Brox, T.: U-net: Convolutional networks for
  biomedical image segmentation. In: Medical Image Computing and
  Computer-Assisted Intervention--MICCAI 2015: 18th International Conference,
  Munich, Germany, October 5-9, 2015, Proceedings, Part III 18. pp. 234--241.
  Springer (2015)

\bibitem{rousseau2021post}
Rousseau, A.J., Becker, T., Bertels, J., Blaschko, M.B., Valkenborg, D.: Post
  training uncertainty calibration of deep networks for medical image
  segmentation. In: 2021 IEEE 18th International Symposium on Biomedical
  Imaging (ISBI). pp. 1052--1056. IEEE (2021)

\bibitem{warfield2004simultaneous}
Warfield, S.K., Zou, K.H., Wells, W.M.: Simultaneous truth and performance
  level estimation (staple): an algorithm for the validation of image
  segmentation. IEEE transactions on medical imaging  \textbf{23}(7),  903--921
  (2004)

\bibitem{wein2015automatic}
Wein, W., Karamalis, A., Baumgartner, A., Navab, N.: Automatic bone detection
  and soft tissue aware ultrasound--ct registration for computer-aided
  orthopedic surgery. International journal of computer assisted radiology and
  surgery  \textbf{10},  971--979 (2015)

\bibitem{wenger2020non}
Wenger, J., Kjellstr{\"o}m, H., Triebel, R.: Non-parametric calibration for
  classification. In: International Conference on Artificial Intelligence and
  Statistics. pp. 178--190. PMLR (2020)

\bibitem{xu2019deepatlas}
Xu, Z., Niethammer, M.: Deepatlas: Joint semi-supervised learning of image
  registration and segmentation. In: Medical Image Computing and Computer
  Assisted Intervention--MICCAI 2019: 22nd International Conference, Shenzhen,
  China, October 13--17, 2019, Proceedings, Part II 22. pp. 420--429. Springer
  (2019)

\end{thebibliography}

\end{document}


\title{Can ultrasound confidence maps predict sonographers' labeling variability?}

\author{
Vanessa Gonzalez Duque\textsuperscript{1,2,*} \and
Leonhard Zirus\textsuperscript{1,*} \and
Yordanka Velikova\inst{1} \and
Nassir Navab\inst{1, 3} \and
Diana Mateus\inst{2}
}

\institute{Computer Aided Medical Procedures, Technical University of Munich, Germany. \\ 
 \and
 LS2N laboratory at Ecole Centrale Nantes, UMR CNRS 6004, Nantes, France\\
 \and
 Computer Aided Medical Procedures, John Hopkins University, Baltimore, USA. \email{vanessag.duque@tum.de} \\
 }
 
 \maketitle 

\section{Supplementary Material}
\begin{table}[h!]
    \centering
    \begin{tabular}{@{}lc@{\hskip 4mm}c@{\hskip 4mm}c@{\hskip 4mm}c@{\hskip 4mm}c@{\hskip 4mm}c@{}}
    \noalign{\smallskip}\hline\noalign{\smallskip}
        $\alpha, \beta$  &Network &  \multicolumn{1}{l}{DSC} &  \multicolumn{1}{l}{IoU} & \multicolumn{1}{l}{Precision} & \multicolumn{1}{l}{Miss rate} & \multicolumn{1}{l}{ASD[mm]} \\
        \noalign{\smallskip}\hline\noalign{\smallskip}
        100, 0.5   &\underline{2ch-D} &0,84$\pm$0,03&  0,75$\pm$0,04	&0,91$\pm$0,04	&0,20$\pm$0,03	&\underline{4,41}$\pm$5,13\\
        400, 0.2   &2ch-D & 0,84$\pm$0,03& 0,75$\pm$0,05	&0,88$\pm$0,06	&0,14$\pm$0,03	&11,87$\pm$5,31\\
        100, 0.2   &2ch-D & 0,83$\pm$0,02& 0,73$\pm$0,03	&0,84$\pm$0,04	&0,15$\pm$0,03	&14,84$\pm$1,15\\
            
        100, 0.5   &1ch-CEconf & 0,83$\pm$0,04& 0,73$\pm$0,06	&0,87$\pm$0,05	&0,17$\pm$0,04	&15,57$\pm$2,68\\
        400, 0.5   &\underline{1ch-CEconf} &0,82$\pm$0,03& 0,72$\pm$0,05	&0,90$\pm$0,06	&0,20$\pm$0,02	&\underline{10,15}$\pm$2,55\\
        100, 0.2   &1ch-CEconf & 0,82$\pm$0,04& 0,72$\pm$0,07	&0,86$\pm$0,06	&0,16$\pm$0,03	&13,58$\pm$4,67\\
        \noalign{\smallskip}\hline\noalign{\smallskip}
    \end{tabular}
    \caption{Metrics comparison of the baseline (U-Net) network for two main uses of the confidence maps: as a second channel (2Ch) and in the loss (CEconf).}
    \label{tab:Ablations}
\end{table}
\begin{figure}[h!]
 	\centering
 	\includegraphics[scale=0.27]{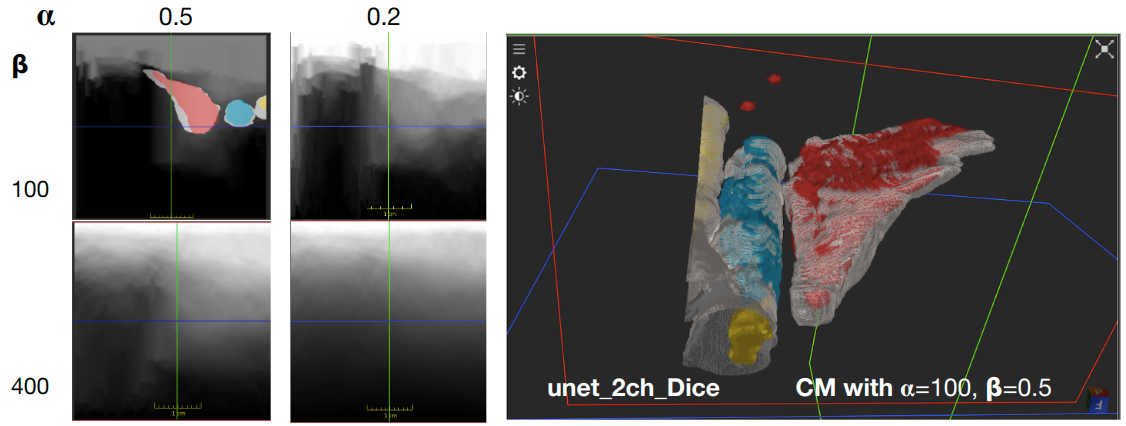}
    \caption{At left, a qualitative view of the confidence maps with different hyper-parameters $\alpha$ and $\beta$ (resolution and adaptativeness). At the right is a 3D view of the inference of Unet-2ch-dice that uses the confidence maps as the second channel, decreasing the appearance of islands.}
    \label{CMversions}
\end{figure}

\begin{figure}[!h]
\includegraphics[width=1.0\linewidth]{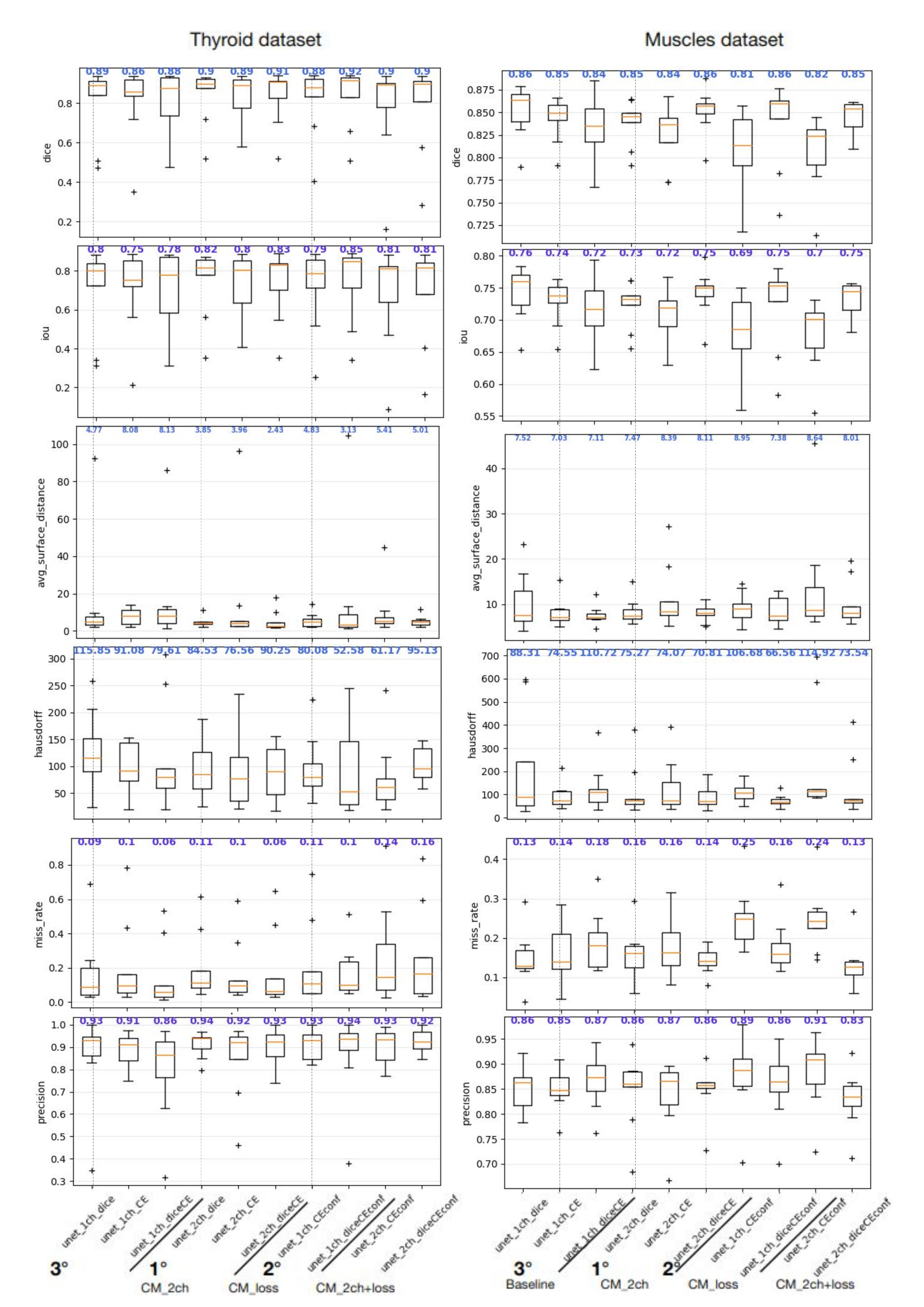}
\caption{First and second columns are thyroid and Muscle metrics respectively. We present results for DSC, IoU, ASD, HD, miss rate and precision with their standard deviation. The three best three configurations for all metrics together are signalized.}
\label{fig1}
\end{figure}